\documentclass[pre,aps,twocolumn]{revtex4-2}
\usepackage{amsfonts}
\usepackage{times}
\usepackage{bbm}
\usepackage{amsmath, amsthm}
\usepackage{graphicx}
\usepackage[normalem]{ulem}
\usepackage{amssymb, algorithm, algpseudocode}
\algnewcommand\algorithmicforeach{\textbf{for each}}
\algdef{S}[FOR]{ForEach}[1]{\algorithmicforeach\ #1\ \algorithmicdo}
\usepackage[title]{appendix}
\usepackage{xcolor,comment}

\begin{document}

\title{Optical Thermodynamics Beyond the Weak Nonlinearity Limit}

\author{Emily Kabat}
\affiliation{Wave Transport in Complex Systems Lab, Physics Department,
Wesleyan University, Middletown, CT-06459, USA}

\affiliation{Department of Applied Physics, Yale University, New Haven, Connecticut 06520, USA}

\author{Shrohan Mohapatra}
\affiliation{ Department of Mathematics and Statistics, University of Massachusetts Amherst, Amherst, 01003-4515, MA, USA}%

\affiliation{Department of Physics, University of Massachusetts Amherst, Amherst, 01003, MA, USA}

\author{P.G. Kevrekidis}
\affiliation{ Department of Mathematics and Statistics, University of Massachusetts Amherst, Amherst, 01003-4515, MA, USA}%

\affiliation{Department of Physics, University of Massachusetts Amherst, Amherst, 01003, MA, USA}

\author{Tsampikos Kottos}
\affiliation{Wave Transport in Complex Systems Lab, Physics Department,
Wesleyan University, Middletown, CT-06459, USA}
\date{\today}

\begin{abstract}
Optical thermodynamics has recently emerged as a theoretical framework describing a Rayleigh-Jeans (RJ) modal power distribution of multimoded nonlinear photonic circuits. However, its applicability is constrained to systems exhibiting weak nonlinear mode-mode interactions. Here, by employing a Transfer Integral Operator, we circumvent this limitation and establish a steady-state \emph{interacting} RJ modal distribution ---referred to as non-ideal RJ (NIRJ)--- with renormalized temperature and optical chemical potential. This also builds a natural bridge with earlier work on grand-canonical 
statistical-mechanical formulations of discrete nonlinear systems.
The theory derives the optical analogue of the compressibility factor, which controls the transition from an ideal, non-interacting equation of state (EoS) to a van der Waals-like interacting EoS.  
\end{abstract}

\maketitle

{\it Introduction --} Multimoded nonlinear photonic systems have been a fertile testbed for the study of many fundamental phenomena, as well as a principal component of many modern technologies. Examples pertaining to the former include foundational questions on wave-turbulence and solitons, optical, BEC and many-body interactions, while on the technological side, one can refer to high-speed communication systems and high-power fiber-based light sources to imaging and cavity microcombs for precision-metrology \cite{DMB2001,FK2005,PH2008,KSVW2010,AGMDP2011,KHD2011,BWF2012,PR2012,SJBRPF2012,RRGBSMEB2012,M2012,MLLF2012,HK2013,RFN2013,PAM2013,RFN2013,RW2013,LBZX2014,WCW2015,PTC2015,AXBRCR2017,WCW2017,XHBAEARC2018,TXWMMM2021,WWCW2022}. In many such frameworks, the intricate nonlinear mode-mode interactions play a crucial role in dictating the light propagation. Though brute-force computations can be invoked for the description of light propagation, they rapidly become power-expensive as the number of modes increases and fail to provide an illuminating description of the underlying beam dynamics. Fortunately, optical thermodynamics (OT) \cite{WHC2019,RFKS2020,MWJC2020,BFKGRMP2020,SKS2021,PSWBWC2022,MWJKCP2023,KRFK2024,ZWHEC2023,COBEP2025} has recently been developed as a new physical framework that circumvents the costly computations while self-consistently describing the complex processes of energy and power exchange among the modes. 

In accordance with the axioms of thermodynamics \cite{LL1980,P2011}, the modal occupation statistics has been derived, following a Rayleigh-Jeans (RJ) distribution defined by two Lagrange multipliers, namely, the optical temperature and chemical potential \cite{WHC2019,RFKS2020,MWJC2020}. These two quantities have been shown to play a central role in controlling the
exchange of power between modes \cite{WHC2019,SKS2021}. Accordingly, an equation of states (EoS) has been derived, while various thermal effects, such as Joule-Thomson photon gas expansion \cite{WJPKC2020,KPASBWRJKSHC2025}, optical phase transitions \cite{RFKS2020}, and negative temperatures \cite{MWJKCP2023}, have been demonstrated. 

In fact, the OT theory parallels the ideal gas formulation of traditional statistical mechanics, where particle interactions are assumed to be very weak. In the optical framework, where the linear supermodes play the role of particles, the mode-mode interactions are facilitated by similarly weak optical nonlinearities. The underlying assumption in both cases is that,  although such weak interactions are important for inducing an ergodicity driving thermalization, the thermal state remains agnostic to the specific nature of the interactions.

At the same time, traditional statistical mechanics has demonstrated that when strong particle-particle interactions are taken into account, new physical phenomena emerge. Examples include the van der Waals and virial corrections to the equation of state, phase transitions, screening, and many-body renormalization, etc \cite{LL1980,P2011}. 

\begin{figure}
    \includegraphics[width=0.5\textwidth]{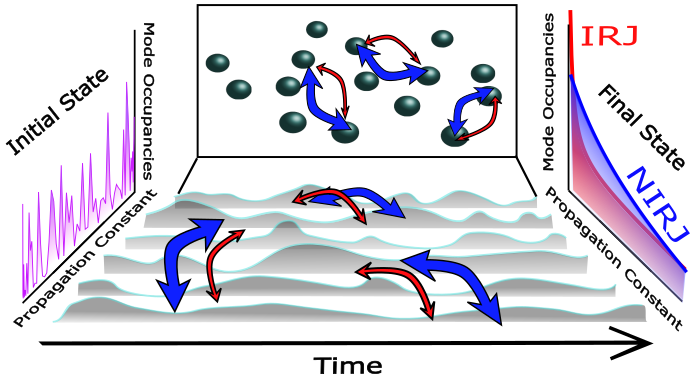}
    \caption{An initial preparation (magenta) evolves toward a final state according to the strength of its nonlinear mode-mode interactions. When these interactions are weak, the system converges to the IRJ distribution predicted by ideal OT (red line), but when the interactions are nonnegligible, it converges to an NIRJ distribution (blue) that deviates from the IRJ. These interactions can be conceptualized through a wave picture (bottom panel), where the modes are exchanging power, or through a particle picture (upper panel), where the different modes undergo billiard-like collisions. }
    \label{schematic}
\end{figure}

Inspired by the phenomena emerging beyond the realm of weak interactions, 
our aim herein is to develop a non-ideal optical thermodynamics which goes beyond the weak nonlinearity domain. Our approach combines the mathematical formulation of 
a Transfer Integral Operator (TIO) scheme \cite{RCKG2000} with a mean-field description of the non-linear interactions, enabling us to establish a methodology that predicts the modal power redistribution of an initial beam excitation in multimoded nonlinear photonic circuits. We find that the interaction-informed thermal state is described by the RJ distribution with optical temperature and chemical potential that are determined by the asymptotic (rather than the initial) value of the linear component of the internal energy, which serves as a grand canonical constant of motion that constrains the linear degrees of freedom (linear supermodes). This centrally important 
notion of {\it asymptotic linear energy} can be very different from its initial value, causing a dramatic deviation of the {\it non-ideal} RJ (NIRJ) from the ideal RJ (IRJ) predicted by an ideal-gas-like OT approach \cite{WHC2019,RFKS2020,MWJC2020} (see Fig. \ref{schematic}). We find that this NIRJ can 
importantly be predicted by (and connected to) the TIO framework to calculate the asymptotic linear energy. Importantly, our scheme predicts the optical analogue of the compressibility factor that controls the transition from an ideal (non-interacting) optical equation of state (EoS) to a van der Waals-like non-ideal EoS (NIEoS) that takes into account the mode-mode interactions. We find that there exists a unique parameter controlling this transition. The latter is given by the ratio between the spectral energy shifts due to background nonlinear interactions and the chemical potential of the equilibrated system. These results offer a bridge between the emerging
realm of OT and the more well-established (in one-dimensional systems) TIO formulation,
while establishing how initial conditions shape the asymptotic features of the system's
thermodynamics and detecting universal trends/dependencies thereof.

{\it Modeling of nonlinear multimoded photonic circuits --} The dynamics of nonlinear multimode photonic networks can be modeled using the framework of time-dependent coupled mode theory. The 
prototypical associated discrete nonlinear Schr{\"o}dinger (DNLS) 
model encompassing diffraction and the cubic nonlinearity
of the Kerr effect~\cite{LEDERER20081,Kevrekidis2009,ILP2012,Rumpf2008}
\begin{equation} \label{DNLS}
i \, \dot{\psi}_k =- J(\psi_{k+1} \, + \, \psi_{k-1}) + \chi |\psi_k|^2 \psi_k ; \, 1 \le k \le N
\end{equation}
where $N$ is the number of nodes (sites) of the network, and $\psi_k$ represents the complex field at each site $k$. Here, we consider only nearest-neighbor couplings with $J=1$, and a Kerr nonlinearity with strength $\chi$. Equation (\ref{DNLS}) describes coupled single-mode resonators, coupled single-mode waveguide arrays, or multicore fibers~\cite{LEDERER20081}. In the former case, the overdot represents a derivative with respect to time $t$ while in the latter two settings it can be thought of as a derivative with respect to the paraxial propagation distance $z$.  

Equation (\ref{DNLS}) is well-known to support two constants of motion~\cite{chriseil,Kevrekidis2009}, namely, the total optical power 
\begin{equation}
\mathcal{A}=\sum_k |\psi_k(t)|^2
\label{norm}
\end{equation}
and the total energy (the electrodynamic momentum flow density in guiding wave settings) 
\begin{equation}
\mathcal{H}=-J\sum_k\left(\psi_k^*\psi_{k+1}+c.c.\right)
+{\chi\over2}\sum_k|\psi_k|^4.
\label{hamiltonian}
\end{equation}
The first term on the RHS of the expression for $\mathcal{H}$, is its linear component $\mathcal{H}_{\rm L}$ while the second term is the nonlinear component $\mathcal{H}_{\rm NL}$. Under the assumption of weak nonlinearity (low power levels, ideal gas), the Hamiltonian is heavily dominated by the linear contribution, i.e., $\mathcal{H}\approxeq \mathcal{H}_{\rm L}$, which often
is used to approximate the relevant conserved quantity in the OT setting~\cite{WJPKC2020}.
Using this ideal OT framework, the Grand Canonical partition function was evaluated using the linear supermode basis (eigenmodes of $\mathcal{H}_{\rm L}$). From there, the equilibrium occupancy of the linear supermodes was 
found to follow the ideal Rayleigh-Jeans (IRJ) distribution 
\begin{equation}
n_\alpha=T_L/(\epsilon_\alpha - \mu_L)
\label{RJdist}
\end{equation}
where $\epsilon_\alpha$ represents the $\alpha$-th linear supermode (eigenvalue of $\mathcal{H}_{\rm L}$).
The Lagrange multipliers $(T_L,\mu_L)$ represent the optical temperature and chemical potential and ensure conservation of the linear energy $\mathcal{H}_L$ and power $\mathcal{A}$ respectively. From the IRJ distribution and the two system invariants $\mathcal{H}_L,\mathcal{A}$ the ideal equation of states (EoS) was derived, $\beta_L\left(h_L-\mu_La\right)=1$ where we have introduced the inverse temperature $\beta_L=1/T_L$, the ensemble-averaged ($\langle ...\rangle$) energy density $h_L=\langle\mathcal{H}_L\rangle/N$ and excitation power $a=\langle\mathcal{A}\rangle/N$.

{\it Beyond the weak nonlinearity approximation --} Obviously, the weak nonlinearity framework of ideal OT collapses when the power levels in the nonlinear multimode arrangements are relatively high, e.g., high peak powers (ultrashort pulses), small modal spacings, long interaction lengths (high-Q resonators). In such cases, an alternative approach that considers the nonlinear component of the Hamiltonian on equal footing is necessary. To this end, we first rewrite the constants of motion using the canonical coordinates $\psi_k=\sqrt{A_k}\exp(i\phi_k)$, as $\mathcal{A}=\sum_k A_k,$ and 
\begin{equation}
\mathcal{H}=-\sum_k2\sqrt{A_kA_{k+1}}\cos(\phi_k-\phi_{k+1})+\frac{\chi}{2}\sum_kA_k^2. 
\label{Hmode}
\end{equation}
In this representation, the Grand Canonical partition function, as was originally
set up in this context in~\cite{RCKG2000}, becomes
\begin{align}\label{PF}
    \mathcal{Z} &= \prod_{n = 1}^{N} \int_{A_n = 0}^{\infty} \mathrm{d}A_n \, \int_{\phi_n = 0}^{2\pi} \mathrm{d}\phi_n \, \operatorname{exp}[-\beta \left(\mathcal{H} - \mu \mathcal{A}\right)]\nonumber\\
    &=(2\pi)^N\int_0^\infty \prod_mdA_mI_0(2\beta\sqrt{A_mA_{m+1}})\nonumber\\
    &\hspace{5mm}\times \exp\bigg\{-\beta\bigg(\frac{\chi}{4}(A_m^2+A_{m+1}^2)-\frac{\mu}{2}(A_m+A_{m+1})\bigg)\bigg\},
\end{align}
where $I_0(\cdot)$ is the zero order modified Bessel function and $\mu,\beta$ are introduced to ensure conservation of $\mathcal{A}$ and (total) $\mathcal{H}=\mathcal{H}_{\rm L}+\mathcal{H}_{\rm NL}$ respectively. As demonstrated in~\cite{RCKG2000}, equation (\ref{PF}) can be exactly solved for any $\mathcal{H}$ and $\mathcal{A}$ by employing the so-called transfer integral operator (TIO) technique
pioneered in~\cite{PhysRevB.6.3409,PhysRevB.11.3535} through the solution of the eigenvalue 
problem
\begin{equation}
\int_{0}^{\infty} \mathrm{d}A_{m} \, \kappa(A_m, A_{m+1}) \, y(A_m)=\lambda y(A_{m+1}),
\label{TIO}
\end{equation}
where the relevant operator is 
\begin{eqnarray}
\kappa(A_m, \, A_{m+1}) = I_0(2 \beta \sqrt{A_m A_{m+1}})
\times 
\nonumber
\\
\operatorname{exp}\bigg(-\frac{\beta \chi }{4} (A_m^2 + A_{m+1}^2) + \frac{\beta \mu}{2}(A_m + A_{m+1})\bigg).
\label{kernel}
\end{eqnarray}
 For multimodal settings where $N\gg 1$, the partition function is dominated by the largest eigenvalue, resulting in $\mathcal{Z}=(2\pi\lambda_0)^N$,  up to exponentially small 
 corrections. Knowledge of $\mathcal{Z}$ allows one to systematically 
 evaluate all other thermodynamic variables. For example, the thermodynamic 
 (grand canonical) potential is $\Omega=-T\ln(\mathcal{Z})$. 
 Similarly, it is well-known in the TIO context~\cite{RCKG2000} (see also~\cite{PhysRevB.6.3409,PhysRevB.11.3535})
 that the average excitation power density and energy density are given by
 \begin{eqnarray}
 a=(1/\beta\lambda_0)\partial\lambda_0/\partial\mu; \hspace{2mm}
 h=-(1/\lambda_0)\partial\lambda_0/\partial\beta+\mu a
 \label{aven}
 \end{eqnarray}
 respectively. Using these expressions, one can derive the NIEoS when nonlinear mode-mode interactions are no longer negligible:
\begin{equation} \label{TIO_EoS}
    \beta(h - \mu a)=f(\beta,\mu,\chi)\,\,{\rm where} \,\, f(\beta,\mu,\chi)=-\frac{\beta}{\lambda_0}\frac{\partial \lambda_0}{\partial \beta}.
\end{equation} 
To some extent, Eq. (\ref{TIO_EoS}) is analogous to the van der Waals EoS of a non-ideal gas, which takes into account the effects of particle-particle interactions. In this respect, $f(\beta,\mu,\chi)$ will be referred to as the compressibility factor.

\begin{figure}
    \includegraphics[width=0.5\textwidth]{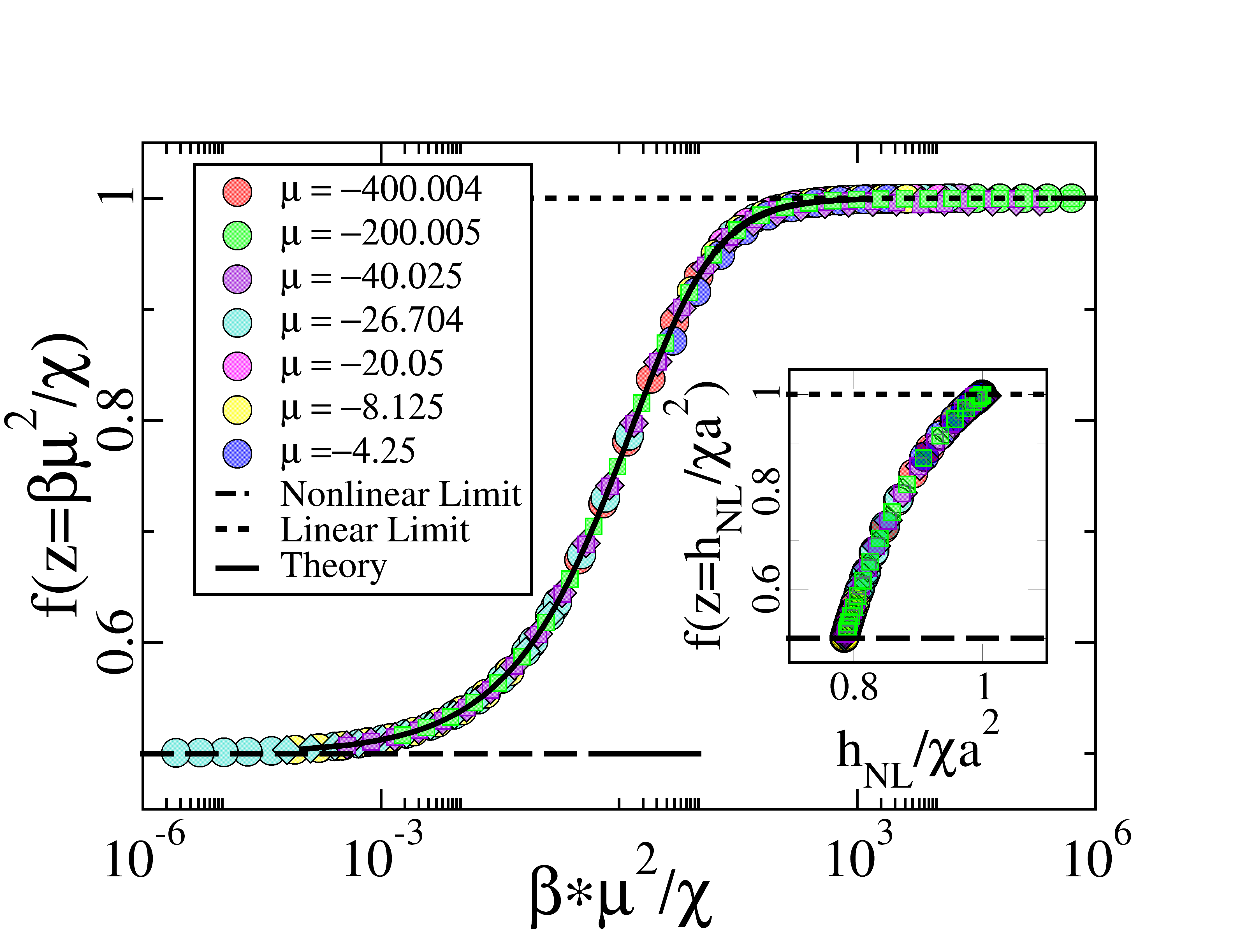}
    \caption{Scaling theory for the difference between the non-ideal equation of state and the OT equation of state. Different colors here represent different TIO chemical potentials, ranging from $\mu=-4.25$ to $\mu=-400$, while different shapes represent different nonlinear strengths (squares represent $\chi=0.1$, circles represent $\chi=1$, and diamonds represent $\chi=10$). In all cases, the equation of state is shown to fall along a single universal curve interpolating between 1/2 (dashed line), representing the strongly nonlinear/equipartition limit, and 1 (dotted line), representing the linear limit in which ideal OT is applicable. We also show a theoretical evaluation of $f(z)$ (solid line). While this equation was evaluated for the case of $z\rightarrow 0$, it is shown to be applicable for the full range of $z$. The same data as shown in the main figure are also reported in terms of the scaling ratio of the extensive variables ${\tilde z} = h_{NL}^\infty/\chi a^2$, which takes values between  ${\tilde z} =\pi/4$ (strong nonlinear interactions, see End Matter) and  ${\tilde z} =1$ (weak nonlinear interactions) \cite{Alba_paper}.}
    \label{scaling}
\end{figure}

{\it Non-ideal Equation of States --} 
The functional form of $f(\beta,\mu,\chi)$ dictates the transition of the NIEoS from the weak nonlinear limit to the nonlinearity-dominant regime. Specifically, we expect that when $\chi\rightarrow 0$, the NIEoS falls back on  the ``ideal gas" scenario, i.e. $f(\beta,\mu,
\chi)\approx 1$. In the other limiting case of sufficiently strong nonlinearities $\mathcal{H}_L\ll\mathcal{H}_{NL}$ the Bessel function in Eq. (\ref{PF}) can be approximated by unity, and the largest TIO eigenvalue can be calculated analytically to be 
\begin{equation}
\lambda_0 \approx \sqrt{\frac{\pi}{2 \beta \chi}} \exp\bigg(\frac{\beta \mu^2}{2 \chi}\bigg) \operatorname{erfc}\Bigg( \sqrt{\frac{\beta\mu^2}{2 \chi}}\Bigg);
\label{lambda0}
\end{equation}
see details in the End Matter.
Consequently, the compressibility factor in the strongly nonlinear regime $f_{NL}$ can be evaluated analytically by a substitution to Eq. (\ref{TIO_EoS}): 
\begin{equation} \label{scaling_theory}
    f_{NL}(\beta,\mu,\chi)=\frac{1}{2}-\frac{z}{2}+\sqrt{\frac{z}{2 \pi}}\frac{e^{-z/2}}{\operatorname{erfc}(\sqrt{z/2})};\, \, z=\beta \mu^2/\chi,
\end{equation}
which for $\chi\rightarrow \infty$ (i.e., $z\rightarrow 0$) gives the other (i.e., highly
nonlinear) limiting value of $f(\beta,\mu,\chi)\rightarrow 1/2$. This can be understood as an embodiment of equipartition in the strongly nonlinear limit, where we assume that each site acts as an uncoupled oscillator, with energy $\sim \chi A_k^2$ (see End Matter). 

It is instructive to notice that the compressibility factor in the nonlinear limit (see Eq. (\ref{scaling_theory})), while in principle depending on $(\beta,\mu,\chi)$, in fact depends 
{\it solely} on the composite variable $z$. Inspired by this observation, we posit that $f(\beta,\mu,\chi)$ is a single-parameter scaling function of $z=\beta\mu^2/\chi$ for all values of $z$. Specifically, we hypothesize the existence of a {\it universal scaling function} that controls the variation of $f(\beta,\mu,\chi)$ as the various parameters $\beta,\mu,\chi$ vary:
\begin{equation}
    \label{CF}
    -\frac{\beta}{\lambda_0}\frac{\partial \lambda_0}{\partial \beta}=f(z={\beta \mu^2\over\chi})=
    \left\{
    \begin{array}{ll}
      1,& z\rightarrow \infty\\
      1/2,& z\rightarrow 0
    \end{array}
    \right.
\end{equation}
Although Eq. (\ref{scaling_theory}) has been derived in the strong nonlinear limit $z\rightarrow 0$, remarkably, our computations suggest that the scaling Eq. (\ref{CF}) applies for the whole range of $z$. In fact, the functional form Eq. (\ref{scaling_theory}) turns out to be valid in all regimes (including the crossover), thus providing a compact summary of the crossover of the compressibility factor from the weak to the strong nonlinear regime. Our numerical results for various $\beta,\mu,\chi$ values are reported in Fig. \ref{scaling} with the corresponding data available at \cite{ZenodoData} and, accordingly, confirm the scaling ansatz Eq. (\ref{CF}) and the applicability of the functional form given by Eq. (\ref{scaling_theory}).

The validity of the one-parameter scaling law Eq. (\ref{CF}) calls for an argument for its explanation. The following heuristic argument provides some intuition regarding the importance of the scaling parameter $z=\beta\mu^2/\chi$. Let us consider the grand-potential density $\omega(a)$. After integrating over microscopic field configurations, the linear kinetic (coupling) term contributes only a constant $\omega_0(T, J)$, since it depends on spatial correlations but not on the invariant $\mathcal{A}$ itself. By contrast, the local Kerr interaction produces a quadratic dependence on $a$, yielding an effective Landau form $\omega(a)\approx\omega_0+{\chi\over 2}a^2-\mu a=\omega_0+{\chi\over 2}(a-a^*)^2-{a^*\mu\over 2}$ where $a^* = \mu/\chi<0$ occurs in the ``unphysical" domain $a<0$, since the applicability of TIO requires $T>0, \mu<0$. Next, we recognize that the Boltzmann weight has the Gaussian form $P(a)\propto \exp\left(-\beta\chi(a-a^*)^2/2\right)$ with the thermal fluctuations of $a$ being controlled by the variance $\sigma_a^2 = 1/(\beta\chi)$. The ratio between $a^*$, where the Gaussian is centered, and $\sigma_a$ defines the scaling parameter $z=(a^*/\sigma_a)^2=\beta \mu^2/\chi$, which \emph{dictates which part of the Boltzmann distribution is accessible to the system, under the physical constraint $a\geq 0$}. When $z\gg1$, the center of the Gaussian is far from the $a=0$ boundary, and the allowed domain samples only its far tail. The distribution is suppressed at large $a\gg 0$ values, so that $P(a)$ is determined by values $a\approx 0$, where $\omega(a)\approx \frac{1}{2}\chi a^2-\mu a\approx |\mu|a$. In this regime, $P(a\geq 0) \propto \exp(-\beta|\mu|a)$ with $\langle a\rangle \sim 1/(\beta|\mu|)$. This is the ideal gas limit, where $h-\mu a\approx |\mu|\langle a\rangle\sim 1/\beta\rightarrow \beta(h-\mu a) = 1$. If instead $z\ll 1$, $a^*$ is less than one thermal width away from the boundary $a=0$, thus cutting the Gaussian close to its center. Here, $\omega(a)\approx \frac{1}{2}\chi a^2$ and $P(a\geq 0)\propto \exp(-\beta \chi a^2/2)$ with $\langle a^2\rangle =1/(\beta\chi)$. This is the interaction-dominated regime where $h-\mu a\approx \frac{1}{2}\chi \langle a^2\rangle =1/(2\beta) 
\rightarrow \beta(h-\mu a)=1/2.$

\begin{figure}
    \includegraphics[width=0.5\textwidth]{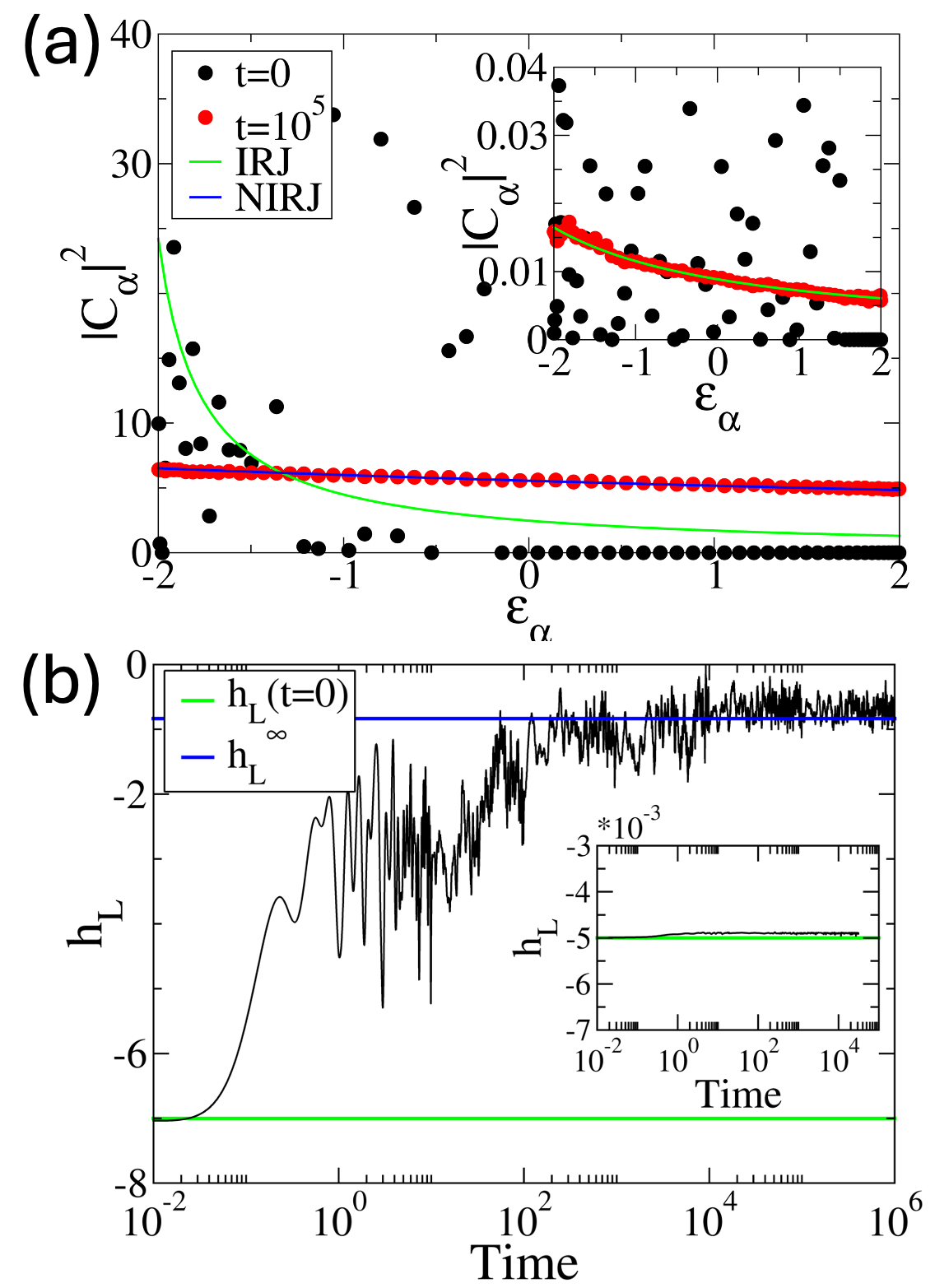}
    \caption{(a) The equilibrium state mode occupancies of a strongly nonlinear system, with total power $a=5.614$. Initial conditions shown as black circles. The IRJ calculated according to the initial linear energy (green line, $\beta=0.18$, $\mu_{\rm eff}=-2.227$) differs dramatically from the distribution at $t=10^5$ (red circles), which is instead predicted by the NIRJ (blue line, $\beta=0.013$, $\mu_{\rm eff}=-13.5$) calculated using the asymptotic linear energy. Inset shows a weakly nonlinear system with $a=0.01$. The IRJ (green line,$\beta = 25.5$ and $\mu_{\rm L}=-4.4$) calculated using $h_L(t=0)=-0.00489$ matches the distribution at $t=10^7$ (red circles). (b) The linear energy of the strongly nonlinear system shown in (a) evolves from its initial value $h_L(t=0)=-7$ (green line) to the TIO prediction for the asymptotic linear energy $h_L^\infty=-0.8361$ (blue line). Inset shows the linear energy of the weakly nonlinear system shown in (a), demonstrating that $h_L=-0.00489$ throughout the evolution. All data is publicly available at \cite{ZenodoData}.}
    \label{equilibrium}
\end{figure}

{\it Non-Ideal Rayleigh-Jeans modal distribution --} When the power levels of the beam are low, the nonlinear contribution $\mathcal{H}_{\rm NL}$ to the total internal energy $\mathcal{H}$ can be neglected. When the nonlinear component $\mathcal{H}_{\rm NL}$ is small enough to be omitted in the calculation of the partition function, the power distribution through
the mode occupancies $|C_\alpha|^2$ across the linear supermodes can be shown to follow an IRJ distribution \cite{WJPKC2020}. The Lagrange multipliers $\mu$ and $T$ characterizing the IRJ distribution are dictated by conservation of $\mathcal{A}$ and $\mathcal{H}_L$. We manifest a typical example of this
scenario in the inset of Fig. \ref{equilibrium}a for an initial excitation $|C_\alpha (t=0)|^2$ (black filled circles) with $a=0.01$, $h_L=-0.005$, and negligible nonlinear energy $h_{NL}=0.0001$. The IRJ 
distribution is reported with a solid green line, while the thermal state evaluated by direct simulations of Eq. (\ref{DNLS}) is reported with red filled circles. Correspondingly, the linear energy (see inset of Fig. \ref{equilibrium}b) is {\it practically} constant over the course of the beam evolution, justifying the underlying assumption that it can be considered as an effective constant of motion. 

This weakly nonlinear scenario has to be contrasted with the case where the nonlinear contribution $\mathcal{H}_{\rm NL}$ is considerable. Here, we still concern ourselves with the power distribution to the ``background modes'', i.e., the extended linear eigenmodes (supermodes) of the linear coupling matrix, which form the delocalized thermal wave gas in equilibrium. We find that the IRJ distribution is {\it incapable} of describing the equilibrium occupation of strongly nonlinear initial excitations. In Fig. \ref{equilibrium}a, we show a typical example of an initial excitation (black filled circles) with $a=5.614$, $h_L=-7$, and $h_{NL}=32.15$. We find that the IRJ prediction (solid green line) deviates dramatically from the final distribution (red filled circles), while the NIRJ (solid blue line) accurately captures this asymptotic distribution.

At the same time, our simulations show that the linear internal energy $h_L$ rapidly deviates from its initial value, see Fig. \ref{equilibrium}b. For longer times, it asymptotically approaches a well-defined equilibrium value (apart from small thermal fluctuations) $\mathcal{H}_L^\infty=\mathcal{H}-\mathcal{H}_{NL}^{\infty}$ that is accurately predicted by the TIO formalism under the constraints of fixed total energy and power $(\mathcal{H}, \mathcal{A})$, respectively. Specifically, it is given by the expectation value associated with the nonlinear energy
in the form:
\begin{align}
    \mathcal{H}_{NL}^{\infty}&=-\frac{(2\pi)^N\chi}{\beta \mathcal{Z}}\int_0^\infty \prod_mdA_m I_0(2\beta\sqrt{A_mA_{m+1}})\nonumber\\
    &\quad\times \frac{\partial}{\partial \chi}\exp\bigg\{-\beta\bigg(\mathcal{H}_{NL}-\mu\mathcal{A})\bigg)\bigg\},
\end{align}
where $\mathcal{H}_{NL}={\chi\over 2}\sum_kA_k^2, \mathcal{A}=\sum_kA_k$. 

It turns out that the single parameter $z$ that dictates the scaling relation Eq. (\ref{equilibrium}) for the compressibility factor can be rewritten in terms of the dimensionless ratio of extensive variables ${\tilde z} = h_{NL}^\infty /\chi a^2 = \sum_n |\psi_n(t\rightarrow \infty) |^4/(\sum_n |\psi_n(t\rightarrow\infty)|)^2$ (see inset of Fig. (\ref{scaling})). It is interesting to point out that the last expression is nothing else than the (normalized) inverse participation number which describes the inverse of spatial occupation of the asymptotic distribution $\{\psi_n(t\rightarrow\infty)\}$ over the available space. The nature of localized thermal states is currently under investigation \cite{Alba_paper}.

These two observations are naturally reconciled by recognizing that the nonlinear energy
renormalizes the energy spectrum, leading to effective mode energies $\epsilon_k\rightarrow \epsilon_k+\Sigma$, where $\Sigma=C \chi a$ and $C$ is a proportionality constant. Alternatively, one can absorb this mean-field nonlinear-induced shift to an effective chemical potential $\mu_{\rm eff}=\mu-\Sigma$. This nonlinear correction is absent from the noninteracting ideal-gas recipe of Ref. \cite{WHC2019}, which assumes unshifted linear dispersions, and therefore cannot reproduce the observed modal occupations within this strongly nonlinear regime. Accordingly, a central
point of our work consists of the recognition 
that within this interacting, nonlinear regime, the linear modes follow a {\it non-ideal} RJ (NIRJ) distribution, 
\begin{equation}
    \label{NIRJ}
    n_\alpha = {T\over \epsilon_\alpha-\mu_{\rm eff}},\quad \mu_{\rm eff}=\mu-\Sigma.
\end{equation}
This, in turn, explains the breakdown of the noninteracting OT framework, as well 
as how to adjust it, when
the assumption $\mathcal{H}_{NL} \ll \mathcal{H}_{L}$ is no longer applicable. The effective chemical potential $\mu_{\rm eff}$ obtained in this way is precisely the chemical potential extracted using the ideal OT methodology when the constraints are the equilibrated $(\mathcal{H}_L^{\infty},\mathcal{A})$. Finally, the proportionality constant $C\approx2$ can be evaluated by imposing the asymptotic linear internal energy constraint using Eq. (\ref{NIRJ}) (see also End Matter).
A representative example which confirms the validity of this scheme is shown in Fig. \ref{equilibrium}a (see green solid line). This demonstrates that the IRJ construction is the correct effective description once the nonlinear background is accounted for via the self-energy. Importantly, also, the temperature and chemical potential
inferred through this process are the {\it same ones} as obtained through the 
---globally applicable within the Gibbsian thermalization regime--- TIO methodology.
This provides a unified (across linearly and nonlinearly dominated such regimes)
perspective of the system's thermodynamics in line with the description of~\cite{RCKG2000}.

Let us finally point out that while $\Sigma$ is perturbatively small at weak nonlinearities so that $\mu_{\rm eff}\approx \mu$, at strong nonlinearities $\Sigma\sim\mathcal{O}(\mu_{\rm eff})$. This latter observation forces us to define another control dimensionless interaction parameter $z_{\rm eff}$, which quantifies the nonlinear interactions at a {\it microscopic} level. Specifically, $z_{\rm eff}$ describes the breakdown of the unshifted ideal RJ prescription for the background linear supermodes. This must be contrasted with the control parameter $z$, which describes transitions from ideal to non-ideal optical gases at the {\it macroscopic} level, as reflected in the NIEoS.

To this end, we first realize that the background modal power occupation $n_\alpha$ of Eq. (\ref{NIRJ}) is controlled by the detuning of the corresponding linear energies $\epsilon_\alpha$ from the chemical potential $\mu_{\rm eff}$. Among them, the lowest-energy $\epsilon_1$ mode has the highest occupation given by $n_1=T/\Delta$ where $\Delta=\epsilon_1-\mu_{\rm eff}$ is the band-edge detuning. On the other hand, the nonlinear interactions introduce a self-energy $\Sigma$ which is at most $\Sigma\sim \chi n_1\sim \chi T/\Delta$. When the nonlinearity-induced self-energy becomes comparable to the band-edge detuning, i.e., $\Sigma\sim \Delta$, the power distribution in the background modes will be altered. We can, therefore ---once again--- identify $z_{\rm eff}\equiv \Delta /\Sigma =\beta\mu_{\rm eff}^2/\chi$ as the dimensionless parameter that controls the microscopic/spectral thermodynamics, i.e., wheather a crossover occurs from an (unshifted) IRJ background ($z_{\rm eff}\gg 1$) to a self-energy-renormalized ($z_{\rm eff}\leq 1$), non-ideal spectral equilibrium distribution of the background supermodes.

{\it Conclusions and Future Challenges --} We have developed an {\it interacting variant}
of Optical Thermodynamics that allows us to derive a van der Waals-like interacting EoS,
accounting for the role of nonlinearity. The optical compressibility factor that dictates the deviations from ideal EoS has been observed to follow a one-parameter (universal) scaling of the temperature, chemical potential, and nonlinearity strength. The macroscopic scaling parameter reflects the relative thermal fluctuations compared to the nonlinearity-induced shift in the minimum of the Landau potential that dictates the Boltzmann weight of the intensity distributions. 
The effect of the nonlinear interactions is also prominent at the microscopic level, where the power distribution among the linear supermodes shows dramatic deviations from the predictions of the ideal (interaction-free) RJ, where appropriate. The deviations of the
interacting (non-ideal) RJ distribution from its ideal counterpart are controlled by a microscopic parameter that is structurally akin to the macroscopic one. They are determined by the ratio of the nonlinearly induced spectral shift to the detuning of a nonlinearly renormalized chemical potential from the spectral band-edge.  This renormalized formulation enables a self-consistent 
connection of the theory with the generically applicable (within the Gibbsian thermalization
regime) transfer integral operator previously developed for such discrete problems, and
thus a unified, systematic treatment thereof.

This methodology allows for an expansion of many thermodynamic analyses into the regime of strong nonlinearities. For example, it becomes possible to design a nonlinear photonic system to exhibit a specific compressibility (i.e., designing how strongly the photonic gas will deviate from the ideal case). Once this target compressibility is chosen, the scaling theory presented here provides a formula for identifying the thermodynamic forces $(\beta, \mu)$, at which point the partition function can be calculated. Using this procedure, our scheme predicts the exact non-ideal RJ 
distribution to which the background modes, at the target compressibility, will thermalize. 
Naturally, this work paves the way for numerous additional veins of research. On the one hand,
it is anticipated that similar perspectives may be applicable to other similar problems,
including, e.g., the Salerno~\cite{salerno_mithun} or the saturable~\cite{saturable} variants of the nonlinear Schr{\"o}dinger
model. Perhaps even more intriguingly, it will be of interest to apply this
interacting RJ framework, beyond the regime of regular Gibbsian thermalization of the DNLS
model. Indeed, preliminary numerical computations suggest that it may describe 
the so-called non-Gibbsian thermalization regime discovered in~\cite{MithunDNLS18}. The study of systems involving disorder would also
be an interesting extension of the present 
study~\cite{Flach2015RandomPotentials}, where it would be less straightforward
to apply TIO-type techniques.
Relevant studies are currently in progress and will be reported in future publications.

\bibliographystyle{unsrt} 
\bibliography{ListOfReferences}{}

@incollection{Flach2015RandomPotentials,
  author    = {Flach, Sergej},
  title     = {Nonlinear Lattice Waves in Random Potentials},
  booktitle = {Nonlinear Optical and Atomic Systems},
  editor    = {Besse, Christophe and Garreau, Jean-Claude},
  series    = {Lecture Notes in Mathematics},
  volume    = {2146},
  pages     = {1--48},
  publisher = {Springer},
  address   = {Cham},
  year      = {2015},
  doi       = {10.1007/978-3-319-19015-0_1}
}

@article{LEDERER20081,
title = {Discrete solitons in optics},
journal = {Physics Reports},
volume = {463},
number = {1},
pages = {1-126},
year = {2008},
issn = {0370-1573},
doi = {https://doi.org/10.1016/j.physrep.2008.04.004},
url = {https://www.sciencedirect.com/science/article/pii/S0370157308001257},
author = {Falk Lederer and George I. Stegeman and Demetri N. Christodoulides and Gaetano Assanto and Moti Segev and Yaron Silberberg}
}

@article{saturable,
  title = {Statistical mechanics of a discrete Schr\"odinger equation with saturable nonlinearity},
  author = {Samuelsen, Mogens R. and Khare, Avinash and Saxena, Avadh and Rasmussen, Kim \O{}.},
  journal = {Phys. Rev. E},
  volume = {87},
  issue = {4},
  pages = {044901},
  numpages = {4},
  year = {2013},
  month = {Apr},
  publisher = {American Physical Society},
  doi = {10.1103/PhysRevE.87.044901},
  url = {https://link.aps.org/doi/10.1103/PhysRevE.87.044901}
}

@article{salerno_mithun,
  title = {Thermalization in the one-dimensional Salerno model lattice},
  author = {Mithun, Thudiyangal and Maluckov, Aleksandra and Manda, Bertin Many and Skokos, Charalampos and Bishop, Alan and Saxena, Avadh and Khare, Avinash and Kevrekidis, Panayotis G.},
  journal = {Phys. Rev. E},
  volume = {103},
  issue = {3},
  pages = {032211},
  numpages = {10},
  year = {2021},
  month = {Mar},
  publisher = {American Physical Society},
  doi = {10.1103/PhysRevE.103.032211},
  url = {https://link.aps.org/doi/10.1103/PhysRevE.103.032211}
}

@article{MithunDNLS18,
  title = {Weakly Nonergodic Dynamics in the Gross-Pitaevskii Lattice},
  author = {Mithun, Thudiyangal and Kati, Yagmur and Danieli, Carlo and Flach, Sergej},
  journal = {Phys. Rev. Lett.},
  volume = {120},
  issue = {18},
  pages = {184101},
  numpages = {6},
  year = {2018},
  month = {May},
  publisher = {American Physical Society},
  doi = {10.1103/PhysRevLett.120.184101},
  url = {https://link.aps.org/doi/10.1103/PhysRevLett.120.184101}
}

@article{PhysRevB.6.3409,
  title = {Statistical Mechanics of One-Dimensional Ginzburg-Landau Fields},
  author = {Scalapino, D. J. and Sears, M. and Ferrell, R. A.},
  journal = {Phys. Rev. B},
  volume = {6},
  issue = {9},
  pages = {3409--3416},
  numpages = {0},
  year = {1972},
  month = {Nov},
  publisher = {American Physical Society},
  doi = {10.1103/PhysRevB.6.3409},
  url = {https://link.aps.org/doi/10.1103/PhysRevB.6.3409}
}

@article{PhysRevB.11.3535,
  title = {Dynamics and statistical mechanics of a one-dimensional model Hamiltonian for structural phase transitions},
  author = {Krumhansl, J. A. and Schrieffer, J. R.},
  journal = {Phys. Rev. B},
  volume = {11},
  issue = {9},
  pages = {3535--3545},
  numpages = {0},
  year = {1975},
  month = {May},
  publisher = {American Physical Society},
  doi = {10.1103/PhysRevB.11.3535},
  url = {https://link.aps.org/doi/10.1103/PhysRevB.11.3535}
}

@Book{Kevrekidis2009,
address = {Heidelberg},
author = {Kevrekidis, P.G.},
edition = {1st},
publisher = {Springer-Verlag},
title = {{The discrete nonlinear Schr{\"o}dinger Equation: Mathematical Analysis, Numerical Computations and Physical Perspectives}},
url = {https://link.springer.com/book/10.1007/978-3-540-89199-4},
year = {2009}
}

@article{chriseil,
  title = {The Discrete Nonlinear Schr\"odinger equation - 20 Years on},
  author = {J. Chris Eilbeck and Magnus Johansson},
  journal = {Localization and Energy Transfer in Nonlinear Systems},
  pages = {pp. 44-67},
  year = {2003}
}

@article{RCKG2000,
  title = {Statistical Mechanics of a Discrete Nonlinear System},
  author = {Rasmussen, K. \O{}. and Cretegny, T. and Kevrekidis, P. G. and Gr\o{}nbech-Jensen, Niels},
  journal = {Phys. Rev. Lett.},
  volume = {84},
  issue = {17},
  pages = {3740--3743},
  numpages = {0},
  year = {2000},
  month = {Apr},
  publisher = {American Physical Society},
  doi = {10.1103/PhysRevLett.84.3740},
  url = {https://link.aps.org/doi/10.1103/PhysRevLett.84.3740}
}

@article{WHC2019,
  title = {Thermodynamic theory of highly multimoded nonlinear optical systems},
  author = {F. O. Wu and E. U. Hassan and D. N. Christodoulides},
  journal = {Nat. Photonics},
  volume = {13},
  pages = {776-782},
  year = {2019},
}

@article{RFN2013,
  author  = {Richardson, D. J. and Fini, J. M. and Nelson, L. E.},
  title   = {Space-division multiplexing in optical fibres},
  journal = {Nature Photonics},
  year    = {2013},
  volume  = {7},
  number  = {5},
  pages   = {354--362},
  doi     = {10.1038/nphoton.2013.94},
  url     = {https://doi.org/10.1038/nphoton.2013.94},
  issn    = {1749-4893}
}

@article{LBZX2014,
author = {Guifang Li and Neng Bai and Ningbo Zhao and Cen Xia},
journal = {Adv. Opt. Photon.},
number = {4},
pages = {413--487},
publisher = {Optica Publishing Group},
title = {Space-division multiplexing: the next frontier in optical communication},
volume = {6},
month = {Dec},
year = {2014},
url = {https://opg.optica.org/aop/abstract.cfm?URI=aop-6-4-413},
doi = {10.1364/AOP.6.000413},
}

@ARTICLE{RRGBSMEB2012,
  author={Ryf, Roland and Randel, Sebastian and Gnauck, Alan H. and Bolle, Cristian and Sierra, Alberto and Mumtaz, Sami and Esmaeelpour, Mina and Burrows, Ellsworth C. and Essiambre, René-Jean and Winzer, Peter J. and Peckham, David W. and McCurdy, Alan H. and Lingle, Robert},
  journal={Journal of Lightwave Technology}, 
  title={Mode-Division Multiplexing Over 96 km of Few-Mode Fiber Using Coherent 6 $\,\times\,$6 MIMO Processing}, 
  year={2012},
  volume={30},
  number={4},
  pages={521-531},
  doi={10.1109/JLT.2011.2174336}}

@article{FK2005,
author = {Shanhui Fan and Joseph M. Kahn},
journal = {Opt. Lett.},
number = {2},
pages = {135--137},
publisher = {Optica Publishing Group},
title = {Principal modes in multimode waveguides},
volume = {30},
month = {Jan},
year = {2005},
url = {https://opg.optica.org/ol/abstract.cfm?URI=ol-30-2-135},
doi = {10.1364/OL.30.000135},
}

@article{XHBAEARC2018,
  author  = {Xiong, Wen and Hsu, Chia Wei and Bromberg, Yaron and Antonio-Lopez, Jose Enrique and Amezcua Correa, Rodrigo and Cao, Hui},
  title   = {Complete polarization control in multimode fibers with polarization and mode coupling},
  journal = {Light: Science \& Applications},
  year    = {2018},
  volume  = {7},
  number  = {1},
  pages   = {54},
  doi     = {10.1038/s41377-018-0047-4},
  url     = {https://doi.org/10.1038/s41377-018-0047-4},
  issn    = {2047-7538}
}

@article{AXBRCR2017,
  title = {Super- and Anti-Principal-Modes in Multimode Waveguides},
  author = {Ambichl, Philipp and Xiong, Wen and Bromberg, Yaron and Redding, Brandon and Cao, Hui and Rotter, Stefan},
  journal = {Phys. Rev. X},
  volume = {7},
  issue = {4},
  pages = {041053},
  numpages = {10},
  year = {2017},
  month = {Nov},
  publisher = {American Physical Society},
  doi = {10.1103/PhysRevX.7.041053},
  url = {https://link.aps.org/doi/10.1103/PhysRevX.7.041053}
}

@article{PH2008,
author = {Francesco Poletti and Peter Horak},
journal = {J. Opt. Soc. Am. B},
number = {10},
pages = {1645--1654},
publisher = {Optica Publishing Group},
title = {Description of ultrashort pulse propagation in multimode optical fibers},
volume = {25},
month = {Oct},
year = {2008},
url = {https://opg.optica.org/josab/abstract.cfm?URI=josab-25-10-1645},
doi = {10.1364/JOSAB.25.001645},
}

@ARTICLE{M2012,
  author={Mafi, Arash},
  journal={Journal of Lightwave Technology}, 
  title={Pulse Propagation in a Short Nonlinear Graded-Index Multimode Optical Fiber}, 
  year={2012},
  volume={30},
  number={17},
  pages={2803-2811},
  doi={10.1109/JLT.2012.2208215}}

@article{PAM2013,
    author = {Pourbeyram, Hamed and Agrawal, Govind P. and Mafi, Arash},
    title = {Stimulated Raman scattering cascade spanning the wavelength range of 523 to 1750 nm using a graded-index multimode optical fiber},
    journal = {Applied Physics Letters},
    volume = {102},
    number = {20},
    pages = {201107},
    year = {2013},
    month = {05},
    issn = {0003-6951},
    doi = {10.1063/1.4807620},
    url = {https://doi.org/10.1063/1.4807620},
    eprint = {https://pubs.aip.org/aip/apl/article-pdf/doi/10.1063/1.4807620/14272004/201107_1_online.pdf},
}

@article{
WCW2017,
author = {Logan G. Wright  and Demetrios N. Christodoulides  and Frank W. Wise },
title = {Spatiotemporal mode-locking in multimode fiber lasers},
journal = {Science},
volume = {358},
number = {6359},
pages = {94-97},
year = {2017},
doi = {10.1126/science.aao0831},
URL = {https://www.science.org/doi/abs/10.1126/science.aao0831},
eprint = {https://www.science.org/doi/pdf/10.1126/science.aao0831},
}

@article{TXWMMM2021,
author = {Mengxi Tan and Xingyuan Xu and Jiayang Wu and Roberto Morandotti and Arnan Mitchell and David J. Moss},
title = {RF and microwave photonic temporal signal processing with Kerr micro-combs},
journal = {Advances in Physics: X},
volume = {6},
number = {1},
pages = {1838946},
year = {2021},
publisher = {Taylor \& Francis},
doi = {10.1080/23746149.2020.1838946},
URL = {   https://doi.org/10.1080/23746149.2020.1838946
},
eprint = {        https://doi.org/10.1080/23746149.2020.1838946
}
}

@article{
KHD2011,
author = {T. J. Kippenberg  and R. Holzwarth  and S. A. Diddams },
title = {Microresonator-Based Optical Frequency Combs},
journal = {Science},
volume = {332},
number = {6029},
pages = {555-559},
year = {2011},
doi = {10.1126/science.1193968},
URL = {https://www.science.org/doi/abs/10.1126/science.1193968},
eprint = {https://www.science.org/doi/pdf/10.1126/science.1193968},
}

@article{WWCW2022,
  title = {Physics of highly multimode nonlinear optical systems},
  author = {L. G. Wright and F. O. Wu and D. N. Christodoulides and F. W. Wise},
  journal = {Nat. Phys},
  volume = {18},
  pages = {1018},
  year = {2022},
}

@article{PR2012,
title = {Condensation of classical optical waves beyond the cubic nonlinear Schrödinger equation},
journal = {Optics Communications},
volume = {285},
number = {24},
pages = {5440-5448},
year = {2012},
issn = {0030-4018},
doi = {https://doi.org/10.1016/j.optcom.2012.07.081},
url = {https://www.sciencedirect.com/science/article/pii/S0030401812007791},
author = {Antonio Picozzi and Sergio Rica},
}

@article{SJBRPF2012,
  author  = {Sun, Can and Jia, Shu and Barsi, Christopher and Rica, Sergio and Picozzi, Antonio and Fleischer, Jason W.},
  title   = {Observation of the kinetic condensation of classical waves},
  journal = {Nature Physics},
  year    = {2012},
  volume  = {8},
  number  = {6},
  pages   = {470--474},
  doi     = {10.1038/nphys2278},
  url     = {https://doi.org/10.1038/nphys2278},
  issn    = {1745-2481}
}

@article{DMB2001,
  title = {Simulations of Bose Fields at Finite Temperature},
  author = {Davis, M. J. and Morgan, S. A. and Burnett, K.},
  journal = {Phys. Rev. Lett.},
  volume = {87},
  issue = {16},
  pages = {160402},
  numpages = {4},
  year = {2001},
  month = {Sep},
  publisher = {American Physical Society},
  doi = {10.1103/PhysRevLett.87.160402},
  url = {https://link.aps.org/doi/10.1103/PhysRevLett.87.160402}
}

@article{KSVW2010,
  author  = {Klaers, Jan and Schmitt, Julian and Vewinger, Frank and Weitz, Martin},
  title   = {Bose--Einstein condensation of photons in an optical microcavity},
  journal = {Nature},
  year    = {2010},
  volume  = {468},
  number  = {7323},
  pages   = {545--548},
  doi     = {10.1038/nature09567},
  url     = {https://doi.org/10.1038/nature09567},
  issn    = {1476-4687}
}

@article{AGMDP2011,
  title = {Condensation and thermalization of classsical optical waves in a waveguide},
  author = {Aschieri, P. and Garnier, J. and Michel, C. and Doya, V. and Picozzi, A.},
  journal = {Phys. Rev. A},
  volume = {83},
  issue = {3},
  pages = {033838},
  numpages = {13},
  year = {2011},
  month = {Mar},
  publisher = {American Physical Society},
  doi = {10.1103/PhysRevA.83.033838},
  url = {https://link.aps.org/doi/10.1103/PhysRevA.83.033838}
}

@article{PSWBWC2022,
  title = {Direct observations of thermalization to a Rayleigh-Jeans distribution in multimode optical fibres},
  author = {H. Pourbeyram and P. Sidorenko and F. O. Wu and N. Bender and L. Wright and D. N. Christodoulides and F. Wise},
  journal = {Nat. Physics},
  volume = {18},
  pages = {685-690},
  year = {2022},
}

@article{WCW2015,
  author  = {Wright, Logan G. and Christodoulides, Demetrios N. and Wise, Frank W.},
  title   = {Controllable spatiotemporal nonlinear effects in multimode fibres},
  journal = {Nature Photonics},
  year    = {2015},
  volume  = {9},
  number  = {5},
  pages   = {306--310},
  doi     = {10.1038/nphoton.2015.61},
  url     = {https://doi.org/10.1038/nphoton.2015.61},
  issn    = {1749-4893}
}

@article{RW2013,
  author  = {Renninger, W. H. and Wise, F. W.},
  title   = {Optical solitons in graded-index multimode fibres},
  journal = {Nature Communications},
  year    = {2013},
  volume  = {4},
  number  = {1},
  pages   = {1719},
  doi     = {10.1038/ncomms2739},
  url     = {https://doi.org/10.1038/ncomms2739},
  issn    = {2041-1723}
}

@article{
MWJKCP2023 ,
author = {A. L. Marques Muniz  and F. O. Wu  and P. S. Jung  and M. Khajavikhan  and D. N. Christodoulides  and U. Peschel },
title = {Observation of photon-photon thermodynamic processes under negative optical temperature conditions},
journal = {Science},
volume = {379},
number = {6636},
pages = {1019-1023},
year = {2023},
doi = {10.1126/science.ade6523},
URL = {https://www.science.org/doi/abs/10.1126/science.ade6523},
eprint = {https://www.science.org/doi/pdf/10.1126/science.ade6523},
}

@article{BWF2012,
  title = {Imaging through nonlinear media via digital holography},
  author = {C. Barsi and W. Wan and J. W. Fleischer},
  journal = {Phys. Rev. E},
  volume = {86},
  pages = {011108},
  year = {2012},
}

@article{PTC2015,
  title = {Seeing through chaos in multimode fibres},
  author = {M. Pl{\"o}schner and T. Tyc and T. Cizm{\'a}r},
  journal = {Nature Photon.},
  volume = {9},
  pages = {529},
  year = {2015},
}

@article{MLLF2012,
  title = {Controlling waves in space and time for imaging and focusing in complex media},
  author = {A. P. Mosk and A. Lagendijk and G. Lerosey and M. Fink},
  journal = {Nature Photon.},
  volume = {6},
  pages = {283},
  year = {2012},
}

@book{HK2013,
  title     = "Mode Coupling and Its Impact on Spatially Multiplexed Systems, Optical Fiber Telectommunications VIB",
  author    = "K.-P. Ho and J. M. Kahn",
  year      = 2013,
  publisher = "Elsevier",
  address   = "New York"
}

@article{ILP2012,
  title = {Nonequilibrium discrete nonlinear Schr{\"o}dinger equation},
  author = {S. Iubini and S. Lepri and A. Politi},
  journal = {Phys. Rev. E},
  volume = {86},
  pages = {011108},
  year = {2012},
}

@article{RFKS2020,
  title = {Optical phase transitions in photonic networks: A spin-system formulation},
  author = {A. Y. Ramos and L. Fern{\'a}ndez-Alc{\'a}zar and T. Kottos and B. Shapiro},
  journal = {Phys. Rev. X},
  volume = {10},
  pages = {031024},
  year = {2020},
}

@article{KRFK2024,
  title = {Nonlinear defect theory of thermal relaxation in complex multimoded systems},
  author = {Kabat, Emily and Ramos, Alba Y. and Fern\'andez-Alc\'azar, Lucas J. and Kottos, Tsampikos},
  journal = {Phys. Rev. Res.},
  volume = {6},
  issue = {3},
  pages = {033114},
  numpages = {8},
  year = {2024},
  month = {Jul},
  publisher = {American Physical Society}
}

@article{MWJC2020,
  title = {Statistical mechanics of weakly nonlinear optical multimode gases},
  author = {K. G. Makris and F. O. Wu and P. S. Jung and D. N. Christodoulides},
  journal = {Opt. Lett.},
  volume = {45},
  pages = {1651},
  year = {2020},
}

@article{SKS2021,
  title = {Controlling optical beam thermalization via band-gap engineering},
  author = {C. Shi and T. Kottos and B. Shapiro},
  journal = {Phys. Rev. Res.},
  volume = {3},
  pages = {033219},
  year = {2021},
}

@article{BFKGRMP2020,
  title = {Classical Rayleigh-Jeans condensation of light waves: observation and thermodynamic characterization},
  author = {K. Baudin and A. Fusaro and K. Krupa and J. Garnier and S. Rica and G. Millot and A. Picozzi},
  journal = {Phys. Rev. Lett},
  volume = {125},
  pages = {244101},
  year = {2020},
}

@book{P2011,
  title     = "Statistical Mechanics",
  author    = "R. K. Pathria and Paul D. Beale",
  year      = 2011,
  publisher = "Elsevier",
  address   = "New York"
}

@book{LL1980,
  title     = "Statistical Physics",
  author    = "L. D. Landau and E. M. Lifshitz",
  year      = 1980,
  publisher = "Butterworth-Heinemann",
  address   = "Oxford"
}

@article{KPASBWRJKSHC2025,
  author  = {Kirsch, Marco S. and
             Pyrialakos, Georgios G. and
             Altenkirch, Richard and
             Selim, Mahmoud A. and
             Beck, Julius and
             Wolterink, Tom A. W. and
             Ren, Huizhong and
             Jung, Pawel S. and
             Khajavikhan, Mercedeh and
             Szameit, Alexander and
             Heinrich, Matthias and
             Christodoulides, Demetrios N.},
  title   = {Observation of {J}oule--{T}homson photon-gas expansion},
  journal = {Nature Physics},
  year    = {2025},
  volume  = {21},
  number  = {2},
  pages   = {214--220},
  doi     = {10.1038/s41567-024-02736-1},
  url     = {https://doi.org/10.1038/s41567-024-02736-1},
  issn    = {1745-2481},
}

@article{WJPKC2020,
  author  = {Wu, Fan O. and
             Jung, Pawel S. and
             Parto, Midya and
             Khajavikhan, Mercedeh and
             Christodoulides, Demetrios N.},
  title   = {Entropic thermodynamics of nonlinear photonic chain networks},
  journal = {Communications Physics},
  year    = {2020},
  volume  = {3},
  number  = {1},
  pages   = {216},
  doi     = {10.1038/s42005-020-00484-1},
  url     = {https://doi.org/10.1038/s42005-020-00484-1},
  issn    = {2399-3650},
}

@misc{ZenodoData,
  author       = {Kabat, Emily and Mohapatra, Shrohan and Kevrekidis, P.G. and Kottos, Tsampikos},
  title        = {{Data for: Optical Thermodynamics Beyond the Weak Nonlinearity Limit}},
  month        = Jun,
  year         = 2026,
  publisher    = {Zenodo},
  note          = {DOI:10.5281/zenodo.19140207}
}

@misc{COBEP2025,
      title={Thermalization of quantum light induced by classical nonlinear wave dynamics}, 
      author={Fouad Chahrour and Şahin K. Ozdemir and Kurt Busch and Ramy El-Ganainy and Armando Perez-Leija},
      year={2025},
      eprint={2511.15100},
      archivePrefix={arXiv},
      primaryClass={physics.optics},
      url={https://arxiv.org/abs/2511.15100}, 
}

@article{ZWHEC2023,
  author    = {Qi Zhong and Fan O. Wu and Absar U. Hassan and Ramy El-Ganainy and Demetrios N. Christodoulides},
  title     = {Universality of light thermalization in multimoded nonlinear optical systems},
  journal   = {Nature Communications},
  year      = {2023},
  volume    = {14},
  number    = {1},
  pages     = {370},
  doi       = {10.1038/s41467-023-35891-9},
  url       = {https://doi.org/10.1038/s41467-023-35891-9},
  issn      = {2041-1723}
}

@article{Rumpf2008,
  title = {Transition behavior of the discrete nonlinear Schr\"odinger equation},
  author = {Rumpf, Benno},
  journal = {Phys. Rev. E},
  volume = {77},
  issue = {3},
  pages = {036606},
  numpages = {8},
  year = {2008},
  month = {Mar},
  publisher = {American Physical Society},
  doi = {10.1103/PhysRevE.77.036606},
  url = {https://link.aps.org/doi/10.1103/PhysRevE.77.036606}
}

@article{Alba_paper,
  author = {Alba Ramos and Emily Kabat and P.G. Kevrekidis and Tsampikos Kottos},
  title = {In preparation},
  year = {2026},
}


\newpage
\onecolumngrid
\section*{End Matter}
\twocolumngrid

\emph{Theoretical Evaluation of $\lambda_0$--} The TIO operator describing the partition function is given by 
\begin{align}
    \kappa(A_m, \, A_{m+1}) = &I_0(2 \beta \sqrt{A_m A_{m+1}})\times \nonumber\\
    &\operatorname{exp}\bigg(-\frac{\beta \chi }{4} (A_m^2 + A_{m+1}^2)\nonumber\\
    &\hspace{1cm}+\frac{\beta \mu}{2}(A_m + A_{m+1})\bigg).
\end{align}
To treat this operator analytically, we begin by expanding the Bessel function as
\begin{equation}
    I_0(x)\approx 1+\frac{1}{4}x^2+...\nonumber
\end{equation}
in the limit of $x\rightarrow0$. We emphasize that this limit can be interpreted as one of two physical conditions: On the one hand, taking the argument of the Bessel function to zero corresponds to taking the high-temperature limit, $\beta\rightarrow 0$. On the other hand, the Bessel function reflects the contribution of the linear energy to the partition function, so approximating the Bessel function as unity also describes systems with negligible linear energy, i.e., in a strongly nonlinear
regime. 

In this limit, the eigenvalue problem becomes
\begin{align}
    \int_0^{A_m}&{\rm d}A_m \kappa^{(0)}(A_m, A_{m+1})y_0^{(0)}(A_m)\nonumber\\ 
    &= \exp\bigg(\frac{-\beta\chi}{4}A_{m+1}^2+\frac{\beta\mu}{2}A_{m+1}\bigg)\nonumber\\
    &\hspace{5mm}\times\int_0^{A_m}{\rm d}A_m\exp\bigg(\frac{-\beta\chi}{4}A_{m}^2+\frac{\beta\mu}{2}A_{m}\bigg)y_0^{(0)}(A_m)\nonumber\\
    &= \lambda_0 ^{(0)}y^{(0)}(A_{m+1}).
\end{align}

Therefore, $y^{(0)}(A_{m})\propto \exp\bigg(\frac{-\beta\chi}{4}A_{m}^2+\frac{\beta\mu}{2}A_{m}\bigg)$, and finally, with the normalization condition, we conclude that 
\begin{equation}
    y^{(0)}(A_{m})=\frac{1}{\sqrt{\lambda_0^{(0)}}} \exp\bigg(\frac{-\beta\chi}{4}A_{m}^2+\frac{\beta\mu}{2}A_{m}\bigg)
    \label{y0}
\end{equation}
\begin{equation}
    \lambda_0^{(0)}=\sqrt{\frac{\pi}{2\beta\chi}}\exp(\frac{\beta\mu^2}{2\chi}){\rm erfc(\sqrt{\frac{\beta\mu^2}{2\chi}}}).
    \label{lambda0}
\end{equation}

This eigenvalue forms the basis of the scaling theory presented in Fig. 2. 

\setlength{\parskip}{1em}
\emph{Equipartition in the Strongly Nonlinear Limit--} To understand the behavior of the equation of state in the strongly nonlinear limit, we begin by explicitly separating out the linear and nonlinear contributions to the energy:
\begin{equation}
    \beta(h-\mu a) = \beta (h_L+h_{NL}-\mu a).
\end{equation}

\begin{figure}
    \includegraphics[width=0.5\textwidth]{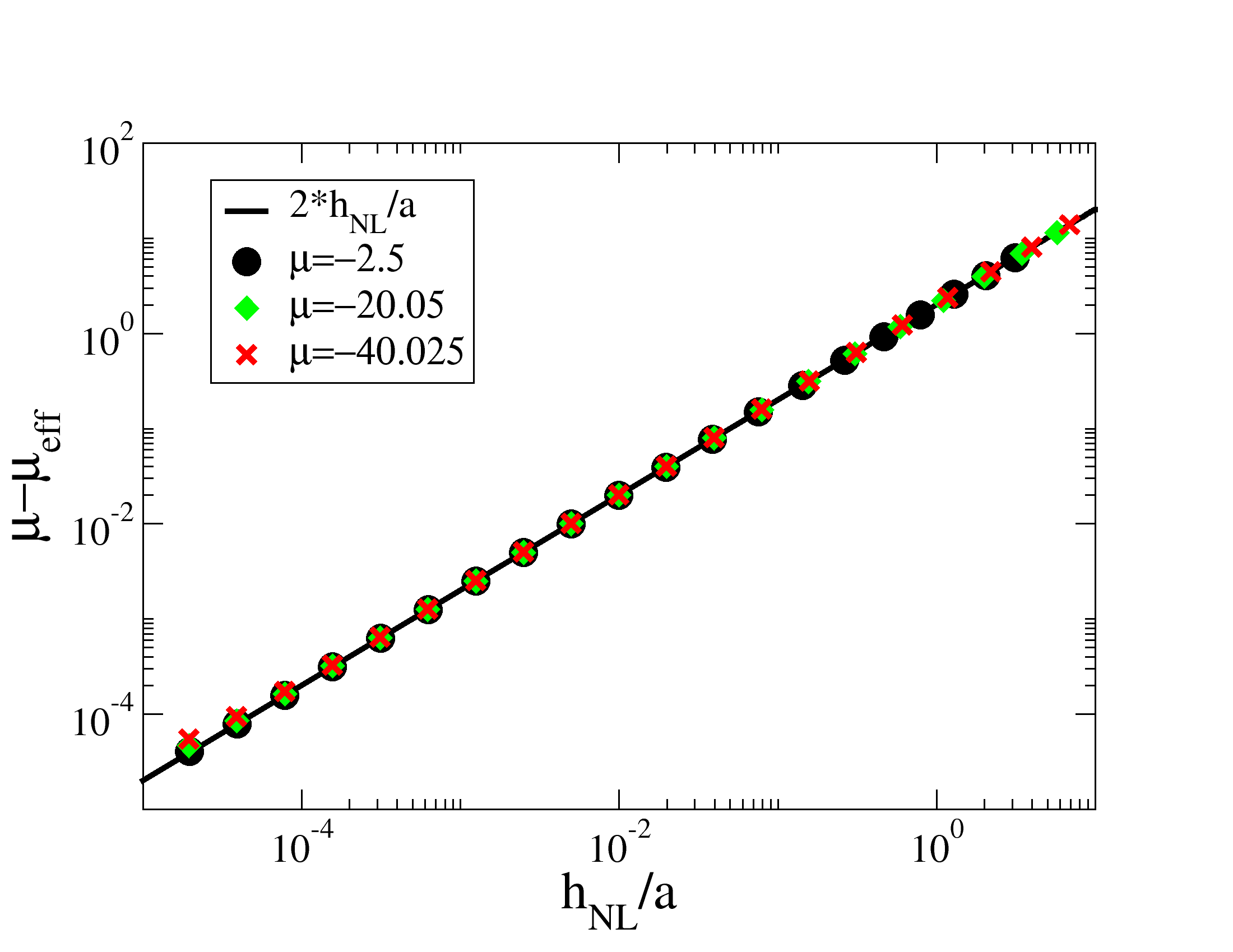}
    \caption{We report the numerical evaluation of $\mu-\mu_{eff}$ as the temperature is varied for chemical potentials $\mu=-2.5$ (black circles), $\mu=-20.05$ (purple circles), and $\mu=-40.025$ (red crosses). In all cases, the shift in the chemical potential falls along the line $\mu-\mu_{eff}=2 h_{NL}/a$. }
    \label{sfig2}
\end{figure}

We then evaluate the chemical potential in terms of the nonlinear energy. Numerically, we have found that $\mu=\mu_{eff}+\frac{2h_{NL}}{a}$ (see Fig. (\ref{sfig2})), which allows for the further simplification 
\begin{align}
    \beta(h-\mu a) &= \beta (h_L+h_{NL}-\mu a)\nonumber\\
    &=\beta\bigg(h_L+h_{NL}-(\mu_{eff}+2h_{NL}/a)\cdot a\bigg)\nonumber\\
    &=\beta(h_L-\mu_{eff} a)-\beta h_{NL}.
\end{align}

Finally, we recognize $\beta(h_L-\mu_{eff} a)=1$ as the ideal EoS, with the result that 
\begin{equation}
    \beta(h-\mu a)=1-\beta h_{NL}.
\end{equation}

This expression allows us to directly parse the role of the nonlinear energy in influencing the non-ideal equation of state: when the nonlinear energy is very small, or the temperature is very high, this expression simplifies to the ideal case, when $\beta(h-\mu a)\approx \beta(h_L-\mu_{eff}a)=1$. However, in systems where the nonlinear energy is non-negligible (and the temperature is finite), the equation of state deviates from the ideal scenario. 

In particular, when the nonlinear energy comes to dominate the total energy so that $h\approx h_{NL}$, we can view the energy through the lens of equipartition. The various sites of the lattice become decoupled, so that each one is effectively an independent oscillator with energy $\sim A_m ^2$. Since the energy has just one degree of freedom corresponding to the amplitude of each site, equipartition is employed with the result that $\beta h_{NL}=\frac{1}{2}$. Therefore, in this limit, 
\begin{equation}
    \beta(h-\mu a)=1-\beta h_{NL}\rightarrow 1-\frac{1}{2}=\frac{1}{2}.
\end{equation}

This evaluation accounts for the saturation of the scaling theory for the equation of state in the limit $z=\beta\mu^2/\chi\rightarrow 0$, which we present as the ``strongly nonlinear'' limit. 

\setlength{\parskip}{1em}
    \emph{Nonlinear Shift to Effective Chemical Potential--} We are concerned with the energy shift to the chemical potential $\mu$ as a result of the nonlinear energy. Written in the basis of the linear supermodes, the total Hamiltonian takes the form 
\begin{equation}
    \mathcal{H}=\mathcal{H}_L+\mathcal{H}_{NL}=\sum_\alpha \varepsilon_\alpha|C_\alpha|^2+\frac{\chi}{2}\sum_{\alpha\beta\gamma\delta} \Gamma_{\alpha\beta\gamma\delta}C_\alpha^*C_\beta^*C_\delta C_\gamma,
\end{equation}
where $\Gamma_{\alpha\beta\gamma\delta}=\sum_m f_\alpha^*(m)f_\beta^*(m)f_\gamma(m)f_\delta(m)=\frac{1}{N^2}\sum_m \exp(i(k_\alpha+k_\beta-k_\gamma-k_\delta)m)\approx \frac{1}{N}\delta(k_\alpha+k_\beta-k_\gamma-k_\delta)$ encodes the four-wave overlap of the supermodes of the linear Hamiltonian at each site $m$ of the lattice. We begin by noting that here, we consider the ordered system of size $N$, where $f_\alpha(m)=\sqrt{2/N}\sin (\varepsilon_\alpha m)$, in which case, $f_\alpha(m)\approx \sqrt{2/N}$. So, we can simplify the above equation to 
\begin{align}
    \mathcal{H}&=\sum_\alpha \varepsilon_\alpha|C_\alpha|^2+\frac{\chi}{2}\sum_{\alpha\beta\gamma\delta} (\sum_m \frac{4}{N^2})C_\alpha^*C_\beta^*C_\delta C_\gamma\nonumber\\
    &=\sum_\alpha \varepsilon_\alpha|C_\alpha|^2+\frac{2\chi}{N}\sum_{\alpha\beta\gamma\delta} C_\alpha^*C_\beta^*C_\delta C_\gamma.
\end{align}

Now, we address the mode amplitudes $C_\alpha$. We can separate the summation over $\alpha, \beta,\gamma,$ and $\delta$ into two distinct cases: combinations such that $\{\alpha, \beta\}=\{\gamma,\delta\}$, which encodes an integrable contribution to the energy, and all combinations for which $\{\alpha, \beta\}\neq\{\gamma,\delta\}$, which will be considered a nonintegrable perturbation. Therefore, the integrable portion of the energy can be written as 
\begin{equation}
    =\sum_\alpha \varepsilon_\alpha|C_\alpha|^2+\frac{2\chi}{N}\sum_{\alpha\beta} |C_\alpha|^2 |C_\beta|^2=\sum_\alpha (\varepsilon_\alpha+2\chi a) |C_\alpha|^2
\end{equation}
where we have used the fact that $\sum_\beta |C_\beta|^2=N\cdot a$. 

In this light, the ideal Rayleigh-Jeans distribution must be modified to accommodate these nonlinear energy shifts. Equivalently, and as presented in the main text, these nonlinear energy shifts can be absorbed into the chemical potential, with the result that 
\begin{equation}
    \mu_{eff}=\mu-2\chi a.
    \label{mu_eff}
\end{equation}
Below we report a figure  showing a numerical evaluation of the constant 2 in Eq. (\ref{mu_eff}). We calculate (using the non-ideal RJ methodology put forth in the main text) the $\mu_{eff}$ associated with a given $(a, h)$ pair, and numerically evaluate the difference $\mu-\mu_{eff}$, finding that this scales nicely as $2\chi a$.

\begin{figure}
    \includegraphics[width=0.5\textwidth]{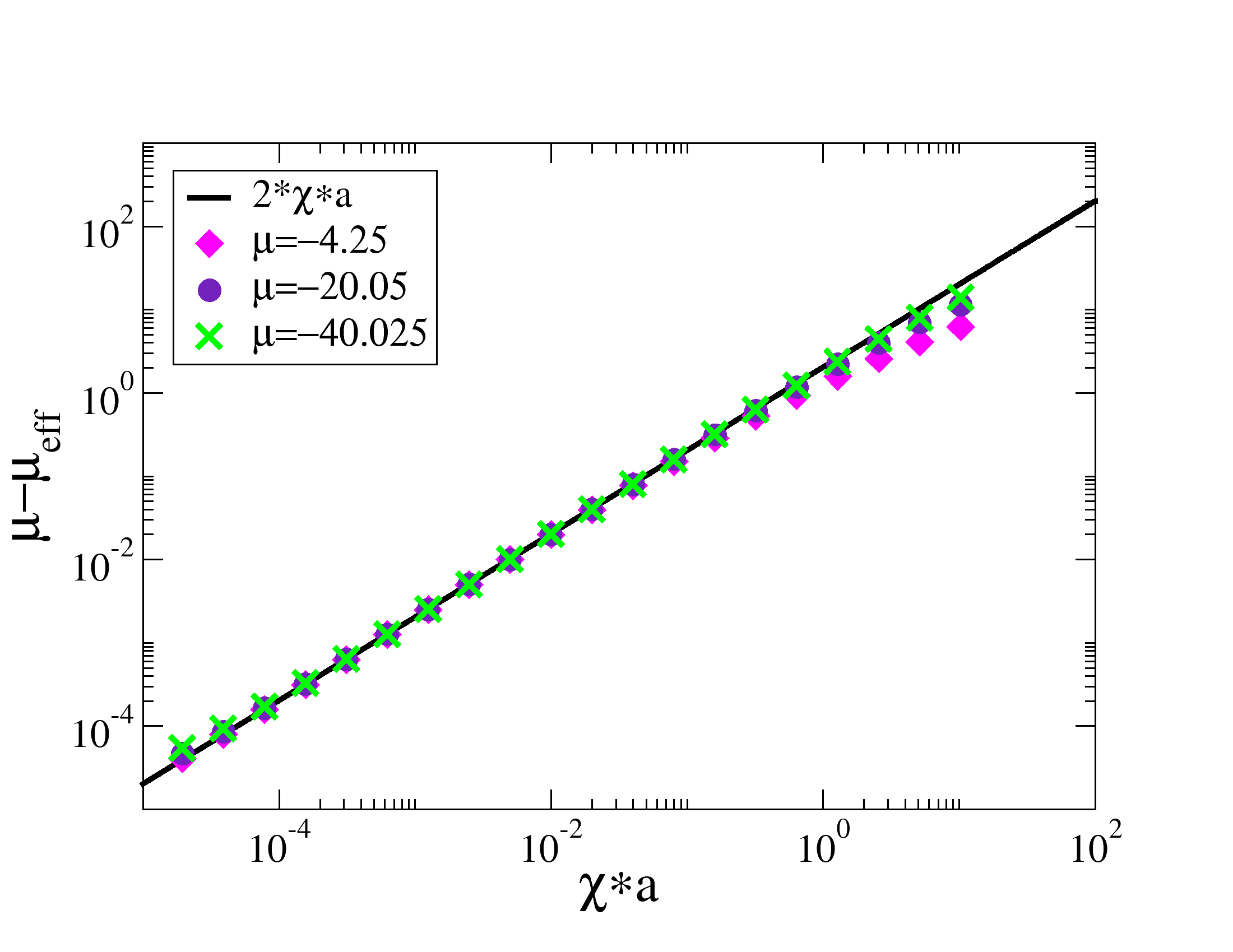}
    \caption{We report the numerical evaluation of $\mu-\mu_{eff}$ as the temperature is varied for chemical potentials $\mu=-4.25$ (magenta diamonds), $\mu=-20.05$ (purple circles), and $\mu=-40.025$ (green crosses). In all cases, the shift in the chemical potential falls along the line $\mu-\mu_{eff}=2\chi a$. }
\end{figure}

\setlength{\parskip}{1em}
\emph{Lower bound of ${\tilde z}$--} We are concerned with bounding the macroscopic scaling parameter ${\tilde z} = \frac{h_{NL}}{\chi a^2} = \frac{1}{2}\frac{\langle A_m^2\rangle}{\langle A_m\rangle^2}$ in the strong nonlinearity limit $\chi\rightarrow\infty$. To this end, we consider the probability distribution of site intensities $A_m$ using $y_0(A_m)^2$, where $y_0(A_m)$ is the dominant TIO eigenvector\cite{RCKG2000}. In the limit of $\chi\rightarrow\infty$, we approximate $y_0(A_m)^2 \rightarrow \sqrt{2\beta \chi/\pi}\exp(-\frac{1}{2}\beta \chi A_m^2)$ (Eq. (\ref{y0}, \ref{lambda0})), where we have taken the limit $\lambda_0 ^2\rightarrow \sqrt{2\beta\chi/\pi}$. In this case, 
\begin{align}
    {\tilde z} &= \frac{1}{2}\frac{\sqrt{2\beta \chi/\pi}\int_0^\infty {\rm d}A_m A_m ^2\exp(-\frac{1}{2}\beta \chi A_m^2)}{\bigg(\sqrt{2\beta \chi/\pi}\int_0^\infty {\rm d}A_m A_m\exp(-\frac{1}{2}\beta \chi A_m^2)\bigg)^2}\nonumber\\
    & = \pi/4.
\end{align}
Therefore, our macroscopic scaling parameter is bounded from below by its strong-nonlinearity limit of ${\tilde z} \rightarrow \pi/4$.

\end{document}